\documentclass[11pt]{article}          
\usepackage[english]{babel}
\makeatletter
\def\@begintheorem#1#2{\trivlist%
 \item[\hskip \labelsep{\sffamily\bfseries #2\ #1}]\itshape}
\newtheorem{teo}{Theorem}[section]
\newtheorem{defi}[teo]{Definition}

\newtheorem{lem}[teo]{Lemma}

\newtheorem{_rem}[teo]{Remark}
\newtheorem{_eje}[teo]{Example}

\makeatother

\newenvironment{beweis}{{\em Proof:}}{\hfill $\rule{2mm}{2mm}$
\vspace{3mm}

}

\DeclareMathAlphabet{\Ma}{U}{msa}{m}{n}
\DeclareMathAlphabet{\Mb}{U}{msb}{m}{n}
\DeclareMathAlphabet{\Meuf}{U}{euf}{m}{n}

\def\got#1{\Meuf{#1}}

\DeclareSymbolFont{ASMa}{U}{msa}{m}{n}
\DeclareSymbolFont{ASMb}{U}{msb}{m}{n}
\DeclareMathSymbol{\hrist}{\mathord}{ASMa}{"16}
\DeclareMathSymbol{\varkappa}{\mathalpha}{ASMb}{"7B}
\DeclareMathSymbol{\CrPr}{\mathord}{ASMb}{"6F}

\newfont{\EinsFont}{cmr7 scaled 1070}
\def\EINS{{\mathchoice{
 \mbox{\unitlength1cm\begin{picture}(.25,.2)\put(0,0){$1$}%
 \put(0.105,0){{\mbox{\fontfamily{cmr}\upshape\small l}}}\end{picture}}}{%
 \mbox{\unitlength1cm\begin{picture}(.25,.2)\put(0,0){$1$}%
 \put(0.105,0){{\mbox{\fontfamily{cmr}\upshape\small l}}}\end{picture}}}{%
 \mbox{\unitlength1cm\begin{picture}(.18,.15)\put(0,0){$\scriptstyle 1$}%
 \put(0.07,0){{\mbox{\fontfamily{cmr}\upshape\EinsFont l}}}\end{picture}}}{%
 \mbox{\unitlength1cm\begin{picture}(.18,.15)\put(0,0){$\scriptstyle 1$}%
 \put(0.07,0){{\mbox{\fontfamily{cmr}\upshape\EinsFont l}}}\end{picture}}}}}

\def\restriction{{\mathchoice{
 \mbox{\unitlength1cm\begin{picture}(.2,.4)%
  \bezier{5}(.07,.3)(.1,.27)(.13,.24)%
  \put(.07,.35){\line(0,-1){.5}}\end{picture}}}{
 \mbox{\unitlength1cm\begin{picture}(.2,.4)%
  \bezier{5}(.07,.3)(.1,.27)(.13,.24)%
  \put(.07,.35){\line(0,-1){.5}}\end{picture}}}{
  \hrist}{\hrist}}}

  \def\al #1.{{\mathcal{#1}}}
  \def\ot #1.{{\got{#1}}}

  \def\ccr #1,#2.{\overline{\Delta(#1,\,#2)}}

  \def\b #1.{{\bf #1}}
  \def\cross#1.{\mathrel{\mathop{\times}\limits_{#1}}}
  \def\C{\Mb{C}}

  \def\R{\Mb{R}}
  
  \def\T{\Mb{T}}
  \def\un{\EINS}
  
\def\rn#1{{\romannumeral#1}}
\def\s #1.{_{\smash{\lower2pt\hbox{\mathsurround=0pt $\scriptstyle #1$}}
\mathsurround=5pt}}
  
\def\set #1,#2.{\left\{\,#1\;\bigm|\;#2\,\right\}}

  \def\wh{\widehat}

  \def\cross #1.{\mathrel{\raise 3pt\hbox{$\mathop\times\limits_{#1}$}}}
\def\b #1.{{\bf #1}}
\def\ab{\allowbreak}

\def\aut{{\rm Aut}\,}
\def\chop{\hfill\break}

\title{\bf Amenability of the Gauge Group.}

\author{
 {\sc Alan Carey}                \\[1mm]
 {\footnotesize Mathematical Sciences Institute,}     \\
 {\footnotesize  Bldg 27, Australian National University,}       \\
 {\footnotesize ACT 0200, Australia.}                             \\ 
 {\footnotesize acarey@wintermute.anu.edu.au}                  \\ 
 {\footnotesize FAX: +61-2-61250759}   \\
\and
{\sc Hendrik Grundling}                                            \\[1mm]
 {\footnotesize Department of Mathematics,}                         \\
 {\footnotesize University of New South Wales,}                      \\
 {\footnotesize Sydney, NSW 2052, Australia.}                             \\
 {\footnotesize hendrik@maths.unsw.edu.au}                  \\
 {\footnotesize FAX: +61-2-93857123}}

\begin{document}
\maketitle

\begin{abstract}
Let $\al G.$ be one of the local gauge groups $C(X,\, U(n)),\ab\;C^\infty(X,\,U(n)),
\ab\;C(X,\, SU(n))$ or $C^\infty(X,\,SU(n))$ where $X$ is a compact
Riemannian manifold. We observe that $\al G.$ has a nontrivial group 
topology, coarser than its natural topology, w.r.t. which it is amenable,
viz the relative weak topology of $C(X,\,M(n)).$
This topology seems more useful than other known amenable topologies
for $\al G..$
We construct a simple fermionic model containing an
action of $\al G.,$ continuous w.r.t. this amenable topology.
\end{abstract}
\vfill
{\bf Running title:} Amenability of the Gauge Group.\chop
\eject
\section{Introduction}

It is still one of the main open problems of mathematical physics
to consistently ``quantise'' the Yang--Mills gauge field theory.
In fact,
at a more general level, we do not have a mathematically consistent
quantum theory of a fermion interacting with a quantised local gauge potential
(abelian or nonabelian) in four dimensional
space--time,- an important component of the standard model of particle
physics. The classical version of this theory has been 
constructed and studied for some time, as well as semiclassical
models, e.g. of a quantized fermion in a
classical gauge potential.

Nevertheless, whatever form the final version of local
quantum gauge field theory takes, it is plausible to assume that there will
be a unital C*-algebra $\al F.$ corresponding to bounded functions of the smeared quantum fields,
on which there is a pointwise continuous action of a topological
group $\al G.$ (the ``local gauge group'') and 
where the physically relevant states
on $\al F.$
are the $\al G.\hbox{--invariant}$ states, which must include
a physically reasonable vacuum state.
In order to ensure that such invariant states exist, the
question arises as to whether there is a natural topology in
which the local gauge group
$\al G.$ is amenable, and w.r.t. which the action on $\al F.$
is continuous. We answer this question in the affirmative in
this letter.

In classical gauge theories  there is a range of local gauge groups $\al G..$
Initially $\al G.$ is taken to be the group of bundle isomorphisms of a
principal fibre bundle $E$  which 
project down to the trivial automorphism on
the base $M.$  
The 
 base space $M$ is usually space--time $\R^4,$ or its 
 one--point compactification, and here we will always take the latter choice.
The fibre group $G$ of $E$ is a compact Lie group, usually taken to
be $SU(n)$ or a product of unitary groups.
Now $\al G.$ has an equivalent expression
as a set of continuous maps from $E$ to $G$ with pointwise multiplication,
so the local gauge groups are usually considered as such  mapping groups
~\cite{Bleecker}.
If $E$ is trivial, we can in fact consider $\al G.$ as a group
of continuous maps from $M$ to $G.$
We take $\al G.\subseteq C(X,\,SU(n))$
where $X$ is a fixed compact Riemannian manifold.
Since the gauge transformations intertwine with the time evolutions
of the classical fields it acts upon, and these are determined by
differential equations (involving a gauge field), 
the gauge transformations must also be in
some class of differentiable functions. The choices for $\al G.$ which
occur in the literature are $C^k(X,\,SU(n)),\ab\;C^\infty(X,\,SU(n))$
and $H^k(X,\,SU(n))\equiv \hbox{k--Sobolev}$ maps for $k$ sufficiently high
\cite{Alb}.
Now each of these gauge groups comes with its own natural topology,
e.g. $C^\infty(X,\,SU(n))$ has the topology of uniform convergence
w.r.t. the differential seminorms in any chart of $X.$
So, ideally we would like to know whether these gauge groups
with their natural topologies are amenable. Since they are not locally
compact, this is a difficult question.

There is some flexibility in the quantum theory as to what topology 
one chooses for the local gauge group $\al G..$ This is because physics only 
requires that the expectation values of physical observables
are continuous w.r.t. physical transformation groups.
Since $\al G.$ is usually factored out, being a symmetry of
non-observable fields,
it would seem that its topology can be changed at will without 
affecting the physical theory.

\section{Amenability results.}
\label{Results}

Let $\al G.$ be a topological group.
\begin{defi}
A bounded function $f:\al G.\to\C$ is {\bf right uniformly continuous}
if $\lim\limits_{x\to e}\|f_x-f\|=0$ 
where $f_x(y):= f(xy)$ for all $y\in\al G..$
Denote the space of these by ${\rm RUC}(\al G.),$ and note that
$f_x\in{\rm RUC}(\al G.)$ whenever $f\in{\rm RUC}(\al G.).$
A {\bf left invariant mean} is a functional $m\in{\rm RUC}(\al G.)^*$
such that $m(1)=1=\|m\|$ and $m(f_x)=m(f)$ for all $x\in\al G.,$
$f\in{\rm RUC}(\al G.).$
We say $\al G.$ is {\bf amenable} if it has a left invariant mean.
\end{defi}
Since ${\rm RUC}(\al G.)$ is a C*--algebra, the definition of
the left invariant mean implies that $m$ is a positive functional, hence a
state.

Recall that if $\al G.$ is amenable w.r.t. some topology, then it
is amenable w.r.t. all topologies which are coarser than this one
(i.e. have fewer open sets).
Let now $\al G.$ be one of the gauge groups above, i.e.
$C^k(X,\,SU(n)),\ab\;C^\infty(X,\,SU(n))$
or $H^k(X,\,SU(n)).$
When $\al G.$ is abelian, i.e. $n=1,$ then it is 
amenable, since all discrete abelian groups are amenable.
If $n>1,$ then $\al G.$ with the discrete topology is NOT amenable.
This is because it contains the group of constant functions,
which is of course isomorphic to $SU(n)\supset SU(2)$ which contains
the free group of two generators, and this is nonamenable
(cf. Pier, Corr~14.3 \cite{Pier}).
However, there is a nontrivial topology on
$\al G.$ w.r.t. which it is amenable:-
\begin{teo}
\label{Basic1}
Let $\al G.=C(X,\,U(n))$ where $X$ is a compact topological space,
equipped with pointwise multiplication. Equip $\al G.\subset
C(X,\,M(n))$ with the relative weak topology of the C*--algebra
$C(X,\,M(n)),$ i.e. the topology defined by the seminorms $p_\omega,$
$\omega\in C(X,\,M(n))^*$ where $p_\omega(g):={\big|\omega(g)\big|}.$
Then $\al G.$ is a topological group which is amenable.
\end{teo}
\begin{beweis} This is an easy application of Paterson's Theorem~2
in~\cite{Paterson}.
By compactness of $X,$ the C*--algebra $C(X,\,M(n))\cong C(X)\otimes M(n)$
is unital and as it is the (unique) tensor product of a commutative
C*--algebra with a finite dimensional C*--algebra, it is nuclear.
Thus we have fulfilled the hypotheses of Paterson's Theorem~2~\cite{Paterson},
hence the unitary group of $C(X,\,M(n))$ with the relative weak topology
is an amenable topological group. However, the unitary group of
$C(X,\,M(n))$ is just $\al G.=C(X,\,U(n))$ and so we are done.
\end{beweis}

The relative weak topology is coarser than the norm topology (hence
than the smooth topology on the subgroup $C^\infty(X,\,U(n)))$
so we do not know whether $\al G.$ with the norm topology
is amenable. 
\begin{teo}
\label{Subgps}
Let $\al G.=C(X,\,U(n))$ where $X$ is a compact $k\hbox{--manifold,}$ and
$\al G.$ has the relative weak topology as above. Then the following subgroups
of $\al G.$ with the relative topology are amenable:
\begin{itemize}
\item[{\rm (i)}] $C^\infty(X,\,U(n))$
\item[{\rm (ii)}] $C(X,\,SU(n))$
\item[{\rm (iii)}] $C^\infty(X,\,SU(n))\,.$
\end{itemize}
\end{teo}
\begin{beweis} (i)
Now $C^\infty(X,\,U(n))\subset C(X,\,U(n))=\al G.$
is dense with respect to the norm topology, i.e. for each $g\in\al G.$
there is a sequence $\{h_n\}\subset C^\infty(X,\,U(n))$ such that
$h_n$ converges to $g$ in norm. But then $h_n$ also converges to $g$
in the weak topology because the norm topology is finer than the weak topology,
i.e. $C^\infty(X,\,U(n))$ is also dense w.r.t. the weak topology.
Since $\al G.$ is amenable, it follows from Paterson Proposition~1 (p720)
\cite{Paterson} that $C^\infty(X,\,U(n))$ is amenable.\chop
(ii) 
Note that $Z(U(n))=\T\un$ and that $SU(n)$ is a non--simply connected
cover of $U(n)\big/\T.$ In fact by $SU(n)\subset U(n)\to U(n)\big/\T$
we get a continuous surjective homomorphism $SU(n)\to U(n)\big/\T$
with kernel $\set\exp(i2\pi k/n)\cdot\un,k=0,\ldots,\,n-1..$ 
Since the centre $Z(\al G.)=C(X,\,\T)$ is a closed subgroup,
the factoring $\al G.\to\al G.\big/Z(\al G.)$ is a continuous 
homomorphism, and as $\al G.$ is amenable, so is
$\al G.\big/Z(\al G.)=C\left(X,\,U(n)\big/\T\right)=:L$
by a direct adaptation of the argument in the proof of 
Prop.~13.1, p118 of Pier~\cite{Pier}.
The factoring map restricts to the subgroup
$K:=C(X,\, SU(n))\subset\al G.,$ denoted 
$\xi:K\to C\left(X,\,U(n)\big/\T\right)=L$
and we already know that its image $L$ is amenable.
Its kernel
 $H:=Z(\al G.)\cap C(X,\, SU(n))$ is Abelian, hence amenable,
so $\xi$ is a homomorphism with an amenable range and kernel.
It now follows from Lemma~\ref{lem1} below that 
$K=C(X,\, SU(n))$ is amenable.
\chop
(\rn3) This follows from (\rn2) by the same argument as (\rn1).
\end{beweis}
\begin{lem}
\label{lem1}
If $K$ is a topological group, and $H$ is a closed normal
subgroup of $K$ such that both $H$ and $L:=K/H$ are amenable,
then $K$ is amenable.
\end{lem}
\begin{beweis}
For locally compact groups, this is in Prop.~13.4, p119 of Pier~\cite{Pier}
and Prop~1.13 of Paterson~\cite{Pat2}. These proofs do
not use the local compactness, so are valid in our case, 
but as they are a bit condensed we
do here an expanded version.\chop
Let $m_1$ (resp. $m_2)$ be a left invariant mean of $H$ (resp. $L).$
Let $f\in{\rm RUC}(K),$ and for $k\in K$ define
$f_{(H)}^k:=(f_k)\restriction H.$ Now
\begin{eqnarray*}
\left\|\big(f_{(H)}^k\big)_x-f_{(H)}^k\right\|&=&
\sup_{y\in H}\big|f_k(xy)-f_k(y)\big|  \\[2mm]
&=& \sup_{y\in H}\big|f(kxk^{-1}ky)-f(ky)\big|  \\[2mm]
&\leq& \sup_{z\in K}\big|f(kxk^{-1}z)-f(z)\big|  \\[2mm]
&=& \left\|f\s kxk^{-1}.-f\right\|\rightarrow 0
\end{eqnarray*}
as $x\rightarrow e,$ since 
$f\in{\rm RUC}(K).$ Thus $f_{(H)}^k\in{\rm RUC}(H)$ and so we can
define $\varphi(k):=m_1(f_{(H)}^k)\,.$ Then
\[
\varphi(hx)=m_1(f_{(H)}^{hx})=m_1\big((f_{(H)}^x)_h\big)=
m_1(f_{(H)}^x)=\varphi(x)
\]
for all $h\in H\,.$ Thus $\varphi$ is constant on cosets $Hx$
so we can identify it with a bounded function 
$\wh\varphi$ on $K/H=L\,.$ Let $\{x_\nu\}\subset L$ be a net
converging to $e,$ then since $L$ has the factor topology, there
is a convergent net $\{\check{x}_\nu\}\subset K,$ $\check{x}_\nu\to e$
such that $\xi(\check{x}_\nu)=x_\nu\,.$
Now
\begin{eqnarray*}
\left\|\wh\varphi\s x_\nu.-\wh\varphi\right\|&=&
\sup_{z\in L}\big|\wh\varphi(x_\nu z)-\wh\varphi(z)\big|  
\leq \sup_{k\in K}\big|\varphi(\check{x}_\nu k)
-\varphi(k)\big|  \\[2mm]
&=& \sup_{k\in K}\big|m_1\left(f_{(H)}^{\check{x}_\nu k}
-f_{(H)}^{k}\right)\big|  
\leq \sup_{k\in K}\left\|f_{(H)}^{\check{x}_\nu k}-f_{(H)}^{k}\right\| \\[2mm]
&=&\sup_{k\in K}\sup_{h\in H}\left|f\s\check{x}_\nu.(kh)
-f(kh)\right|=\sup_{y\in K}\left|f\s\check{x}_\nu.(y)-f(y)\right| \\[2mm]
&=& \big\|f\s\check{x}_\nu.-f\big\|\rightarrow 0
\end{eqnarray*}
as $\check{x}_\nu\to e\,.$ Thus $\wh\varphi\in{\rm RUC}(L),$ and so we 
can define
$M(f):=m_2(\wh\varphi),$ and by linearity and boundedness we have that
$M$ is a functional on ${\rm RUC}(K).$ For invariance, observe first that
$\big(f_k\big)_{(H)}^x=\big(f_{kx}\big)\restriction H=f_{(H)}^{kx}$
for $k\in K,$ hence
$m_1\big((f_k)_{(H)}^x\big)=m_1(f_{(H)}^{kx})=\varphi(kx)=\varphi_k(x)\,.$
Now for $h\in H$ we have
$\varphi_{hk}(x)=\varphi(hkx)=\varphi(kx)=\varphi_k(x),$
hence the map $(k,\,x)\to\varphi_k(x)$ in the first entry depends only
on the coset $Hk.$ Then it follows that
\[
\varphi_k(hx)=\varphi(khx)=\varphi\big((h^{-1}kh)x\big)
=\varphi\s h^{-1}kh.(x)=\varphi_k(x)\;,
\]
hence $\varphi_k(x)$ is also constant on the coset $Hx$ so it factors
to a map on $L:$
 $\wh{(\varphi_k)}=(\wh\varphi)\s\xi(k)..$ Now we have:
\[
M(f_k)=m_2\big(\wh{(\varphi_k)}\big)=m_2\big((\wh\varphi)\s\xi(k).\big)
=m_2(\wh\varphi)=M(f)\;.
\]
So we obtain an invariant mean $M$ for $K,$ so it is amenable.
\end{beweis}

The real usefulness of Theorem~\ref{Subgps}, 
lies in the following well-known facts:
\begin{teo}
\label{vacuum}
Let 
$\alpha:\al G.\to\aut\al F.$ be a pointwise continuous action of 
a topological group
$\al G.$ on an unital C*--algebra $\al F..$ 
\begin{itemize}
\item[(i)] If $\al G.$ is amenable, there is a 
$\al G.\hbox{--invariant}$ state on $\al F..$
\item[(ii)] If there is a $\al G.\hbox{--invariant}$ state 
$\omega$ on $\al F.,$
then in its GNS--representation $(\pi_\omega,\,\al H._\omega,\,\Omega_\omega),$
there is a strong--operator continuous unitary representation
$U:\al G.\to\al B.(\al H._\omega)$ such that 
$U_g\pi_\omega(A)U_g^{-1}=\pi_\omega(\alpha_g(A))$ for all $A\in\al F.,$
and $U_g\Omega_\omega=\Omega_\omega$ for all $g\in\al G.,$ where
$\Omega_\omega$ is the GNS--cyclic vector.
\end{itemize}
\end{teo}
\begin{beweis}
For completeness, here are the proofs.\chop
(i) Choose a state $\omega$ on $\al F.$ and define for each
$A\in\al F.$ a function
$f^A(g):=\omega\big(\alpha\s g^{-1}.(A)\big)\,,$ $g\in\al G.\,.$
Then $f^A$ is bounded and
\begin{eqnarray*}
\left\|f^A_x-f^A\right\|&=&
\sup_{g\in G}\big|f^A(xg)-f^A(g)\big|  
= \sup_{g\in G}\left|\omega\left(\alpha\s g^{-1}.\big(
\alpha\s x^{-1}.(A)-A\big)\right)
\right|  \\[2mm]
&\leq& \big\|\alpha\s x^{-1}.(A)-A\big\|\rightarrow 0
\end{eqnarray*}
as $x\to e\,.$ Hence $f^A\in{\rm RUC}(\al G.),$ and so we can
define $\varphi(A):=m(f^A)$ where $m$ is the left invariant mean
of $\al G.\,.$ Then $\varphi$ is linear, positive and normalised,
i.e. a state. Observe that 
\[
f^{\alpha_g(A)}(x)=\omega\big(\alpha\s x^{-1}.(\alpha_g(A))\big)
=\omega\left(\alpha\s(g^{-1}x)^{-1}.(A)\right)=f^A(g^{-1}x)=f^A\s g^{-1}.(x)
\]
hence $\varphi(\alpha_g(A))=m(f^A\s g^{-1}.)=m(f^A)=\varphi(A),$
i.e. $\varphi$ is an invariant state.\chop
(ii) Recall that $\al H._\omega$ is the closure of the factor space
$\al F./N_\omega$ w.r.t. the Hilbert norm $\|\xi(A)\|^2:=
\omega(A^*A)$ where $\xi:\al F.\to\al F./N_\omega$ denotes the
factor map, and $N_\omega:=\set A\in\al F., \omega(A^*A)=0.$
is the left kernel of $\omega.$ Then $\pi_\omega(A)\xi(B)=\xi(AB)$
and $\Omega_\omega=\xi(\un)\,.$
Now since $\omega$ is $\al G.\hbox{--invariant},$ $\alpha_g$ preserves
$N_\omega,$ hence $U_g\xi(A):=\xi(\alpha_g(A))$ is well-defined,
and extends to a unitary on $\al H._\omega$ by invariance of $\omega.$
That $g\to U_g$ is a homomorphism is clear, covariance follows from
\[
U_g\pi_\omega(A)U_g^{-1}\xi(B)=\xi\left(\alpha_g\big(A\alpha_{g^{-1}}(B)\big)
\right) = \xi(\alpha_g(A)B)=\pi_\omega(\alpha_g(A))\xi(B)
\]
and $U_g\Omega_\omega=\xi(\alpha_g(\un))=\xi(\un)=\Omega_\omega$ is clear.
As for strong--operator continuity, by the inequality
\begin{eqnarray*}
\left\|U_g\psi-\psi\right\|&=&
\big\|U_g(\psi-\varphi)+U_g\varphi-\varphi+\varphi-\psi\big\|   \\[2mm]
&\leq& 2\|\psi-\varphi\|+\|U_g\varphi-\varphi\|
\end{eqnarray*}
for $\psi\in\al H._\omega,\;\varphi\in\al F./N_\omega,$ it suffices to 
verify strong--operator continuity of $U_g$ on the dense space
$\al F./N_\omega.$ This now follows from
\[
\big\|U_g\xi(A)-\xi(A)\big\|^2=\left\|\xi(\alpha_g(A)-A)\right\|^2
=\omega\big((\alpha_g(A)-A)^*(\alpha_g(A)-A)\big)
\leq\|\alpha_g(A)-A\|^2
\]
and the pointwise norm continuity of the action.
\end{beweis}
Below we will construct a simple model
of a Fermion with a continuous action of the amenable group
$C(X,\,U(n))$ on it.

Note that since the relative weak topology is coarser than the norm topology
on $C(X,\,SU(n)),$ and the smooth topology on $C^\infty(X,\,SU(n)),$
any continuous map w.r.t. the relative weak topology is in fact also continuous
w.r.t. these two topologies. In particular, any continuous action or
unitary representation of the groups in Theorem~\ref{Subgps}
w.r.t. the relative weak topology is also continuous w.r.t. the 
usual topologies.

In physical models of gauge field theory, we have an action of the local gauge group
$\al G.\subseteq C(X,\,SU(n))$ where $X$ is compactified
space--time on a C*-algebra $\al F.,$ as well as an action of the translation 
group $\R^4$ on $\al F.$ which intertwines with the natural action
of translations on $\al G..$ 
Specifically there is a
smooth action $\beta:\R^4\to{\rm Diff}\,X$ of the translation
group, leaving the point at infinity invariant which produces the natural
action of $\R^4$ on $\al G.$ by
\[
\big(\alpha_\b a.(g)\big)(x):=g\big(\beta_\b a.(x)\big)\,,\qquad x\in X,\;\b a.\in\R^4,\; 
g\in\al G.\;.
\]
This action is the restriction of the natural action of $\R^4$ on 
$C(X,\,M(n)),$ hence pointwise continuous w.r.t. the relative weak topology
on $\al G..$
Thus if we combine the actions on $\al F.,$ we have in fact an action
of the semidirect product $K:=\al G.\cross\alpha.\R^4$ on $\al F..$
Now $\al G.$ is an amenable closed normal subgroup of $K,$
and as $K/\al G.=\R^4$ is amenable, it follows from Lemma~\ref{lem1}
that $K$ is amenable.  
Thus, providing that the action of $K$ on $\al F.$ is pointwise continuous,
there will be $K\hbox{--invariant}$ states, hence gauge invariant and translation
invariant vacua.
The relative weak topology on $\al G.$ seems to be easier for construction
of models where the action of $K$ on $\al F.$ is continuous.

Concerning the freedom which we have to adjust the topology of the local gauge group,
if $\al G.$ acts on a quantum field on $X$ in a local way, then in physical applications
we must have covariant unitary representations
$U:\R^4\to\al U.(\al H.),$ $V:\al G.\to\al U.(\al H.)$ with
$U_\b a.V_gU_\b a.^*=V\s\alpha_{\b a.}(g).\;.$
If we require that
$\b a.\to U_\b a.$ is strong operator continuous, then so is the map
$\b a.\to V\s\alpha_{\b a.}(g).\;,$
which suggests that we should at least equip $\al G.$ with the finest
group topology which makes the maps $\b a.\in\R^4\to\alpha_\b a.(g)$
continuous.
However, there is no strong argument why we should require the
strong operator continuity for $\b a.\to U_\b a.$ since at this point the
the system still contains nonphysical degrees of freedom.
Only for the final gauge invariant theory will the requirement of strong operator
continuity for representations of the translations be justified.

\section{Relation to other work.}
\label{Context}

It has been known for some time from work of Baez \cite{Baez}
that there is a Hausdorff group topology, weaker than the $C^\infty\hbox{-topology,}$ in which the
gauge group $C^\infty(X,G)$ is amenable where $G$ is locally compact,
and $X$ is a compact Riemannian $k\hbox{--manifold.}$ However, for the case
where $G$ is compact, Baez's construction results in the amenability
of a weaker group topology that is precompact. The easiest way to see it is to notice that the
`cylinder functions' appearing on p.3 in his ArXiv preprint~\cite{Baez} 
by their very definition, factor through continuous functions on finite products of copies of
$G$, and thus are almost periodic functions, continuous with regard to the Bohr topology on
the gauge group (the finest precompact topology coarser than the original one). But it is a
well-known classical result that the space of almost periodic functions on a topological group
supports a unique bi-invariant mean. 
Moreover this topology is not obviously natural from
the viewpoint of gauge groups acting by automorphisms of operator algebras.

There has been another known weaker amenable group topology on the group $C^\infty(X,G)$,
which is
the topology of convergence in measure. Specifically, 
if the group $G$ is locally compact amenable,
then the
group of measurable maps from the standard Borel space $X$, equipped with a non-atomic
probability measure $\mu,$ to $G$
is {\it extremely amenable,} that is, has a fixed point in every compact space on which it
acts
continuously cf.~Theorem 2.2 in~\cite{Pestov1}.
In the case where $G$ is compact, this is a result obtained a long time ago independently by
Glasner
(published much later in~\cite{Glasner})
and Furstenberg and B. Weiss (unpublished). If $G$ is a compact Lie
group and the 
measure $\mu$ has full support, the topology of convergence
in measure on the group of maps coincides with the $L^p$-topology for any $1\leq p<\infty$. 
Because of extreme amenability of $C(X,G)$ in this topology, this group has no non-trivial
finite
dimensional unitary representations. This is because
 every such a representation is a continuous
homomorphism
to some $U(n)$ and leads to an action of $C(X,G)$ on $U(n)$ without fixed points. Thus, it is
minimally
almost periodic, and incompatible with the Bohr topology.

The relative weak topology on the group $C(X,U(n))$ is more interesting, because it is
actually finer than both the Bohr topology and the topology of convergence in measure. This
can be seen by applying the fact 
(used, for instance, by Paterson in~\cite{Paterson}) that the weak topology on a
C*-algebra $\al A.$ coincides with the ultraweak topology coming from the universal
enveloping von Neumann algebra of $\al A..$ It follows that every finite-dimensional unitary
representation of 
$C(X,U(n))$ is relatively weakly continuous, and therefore the Bohr topology on the unitary
group of $\al A.$ is weaker than the relative weak topology. Further, the von Neumann algebra
$\al M.:=L^\infty(X,M(n))$ is an enveloping von Neumann algebra for $C(X,M(n))$ (under the natural
embedding, where a non-atomic measure $\mu$ on $X$ is presumed fixed), and therefore the restriction
of the embedding to the unitary group of $C(X,M(n))$, that is, $C(X,U(n))$, is continuous with
respect to the relative weak topology on the first group and the ultraweak 
(i.e. $\sigma(\al M.,\al M.^*))$
topology on the second. However, it can be proved that the ultraweak topology on the group of
measurable maps from $X$ to $U(n)$ is the topology of convergence in measure.

In addition, the relative weak topology on $C(X,U(n))$ is not precompact.
To see this, recall that it
coincides with the ultraweak topology coming from the universal 
enveloping von Neumann algebra of $C(X,U(n)),$ and the unitary group of 
a von Neumann algebra $\al M.$ with the ultraweak topology is precompact if 
and only if $\al M.$ is purely atomic.
(For the proof of this, one looks at
maximal abelian von Neumann subalgebras 
and uses the decomposition of commutative Von Neumann algebras
in terms of basic types, as in Sect.~9.4 of~\cite{Kadison}).
 However, this is not the case with the 
universal enveloping von Neumann algebra of $C(X)\otimes M(n).$ 
Therefore, the group $C(X,U(n))$ is not precompact

Giordano and Pestov have pointed out to us that one can prove a stronger fact
about the unitary group $C(X,U(n))$
with the relative weak topology: it is strongly amenable in the sense of
Glasner~\cite{Glasner2}
that is, every proximal continuous action of this group on a
compact space has a fixed point. This observation follows if instead of Paterson's theorem,
used by us in the proof of Theorem 2.2, one applies the stronger Corollary 3.7 
in~\cite{Pestov2}.
They also independently noted the  example in Theorem 2.2
but never published it because they did not find any applications of this fact.

\section{Example.}

We construct a model of a Fermion on space--time with a local
gauge transformation.

Let $X$ be the one--point compactification of space--time, and
let $\mu$ be any measure on $X.$ Consider the Hilbert space
$\al H.:= L^2(X,\mu,\C^n)=L^2(X,\mu)\otimes\C^n$
where $\C^n$ has its usual inner product $\overline{\b z.}\cdot\b z.\,.$
Then there is a representation $\rho:C(X,\,M(n))\to\al B.(\al H.)$
by $\big(\rho(A)\psi\big)(x):=A(x)\psi(x)$ for all 
$A\in C(X,\,M(n)),$ $\psi\in\al H.$ and $x\in X\,.$
Observe that
 $\rho$ restricts to a unitary representation of the group
$\al G.=C(X,\,U(n))$ on $\al H.$ which is clearly continuous
w.r.t. the weak topology.

Let $\al F.(\al H.)$ be the C*--algebra of the canonical
anticommutation relations over $\al H.,$
i.e. the simple C*--algebra generated by the set $\set a(\psi),
{\psi\in\al H.}.$ satisfying the relations
\[
\{a(\psi),\,a(\xi)\}=0\,\qquad
\{a(\psi),\,a(\xi)^*\}=(\psi,\xi)\un\quad\forall\;\psi,\,\xi\in\al H.
\]
where $\{\cdot,\cdot\}$ denotes the anticommutator,
and such that $\psi\to a(\psi)$ is an antilinear map, cf.~\cite{BR}.
Then we obtain an action
$\gamma:\al G.\to\aut\al F.(\al H.)$ by
$\gamma_g(a(\psi)):=a(\rho(g)\psi)$ for $\psi\in\al H.,$ $g\in\al G..$
Now
\[
\|\gamma_g(a(\psi))-a(\psi)\|=\left\|a\big(\rho(g)\psi-\psi\big)\right\|
=\left\|\rho(g)\psi-\psi\right\|
\]
from which it follows that $\gamma:\al G.\to\aut\al F.(\al H.)$
is a pointwise continuous action w.r.t. the relative
weak topology of $\al G..$
Thus by Theorem~\ref{vacuum}, $\al F.(\al H.)$ has a
$\gamma\hbox{--invariant}$ state.
(Of course since the Fock state is clearly $\gamma\hbox{--invariant},$
this is not new).
Note that when $\mu$ is absolutely continuous w.r.t.
the Lebesgue measure, there is a unitary representation
of the translation group $\R^4$ on $\al H., $ and this produces
an action of $\R^4$ on $\al F.(\al H.)$ which intertwines with the action of 
$\R^4$ on $\al G..$

A natural way to extend this trivial example, is to add a classical gauge field
by tensoring on to $\al F.(\al H.)$ the C*--algebra $C_b(\al C.),$ i.e.
the algebra of bounded continuous functions on the space of connections
$\al C.$ of a trivial principal $SU(n)\hbox{--bundle}$ over $X,$ where $\al C.$
is equipped with its natural topology. At this point we encounter a problem, 
in that the action of the gauge group $\al G.:=C^\infty(X,SU(n))$
on $\al C.$ is continuous w.r.t. the smooth topology, but not w.r.t. the
relative weak topology. The way to solve this is to observe that only the 
topology of the orbit space $\al C./\al G.$ has physical significance,
so we can change the topology on the orbits of $\al G.$ on $\al C..$
In particular we can change the topology to make the maps
$g\in\al G.\to g\cdot A\in\al C.$ continuous w.r.t. the
relative weak topology on $\al G.,$ where $A$ ranges over $\al C..$
Thus one obtains an action of $\al G.$ on $C_b(\al C.)$
(note that $C_b(\al C.)$ changes with the topology on $\al C.)$ 
which is continuous
w.r.t. the relative weak topology, and combines with the
action $\gamma$ above to give a continuous action on the
tensor product.

\section*{Acknowledgements.}

We are very grateful to Prof. V. Pestov (University of Ottawa) who
provided us with the proper mathematical context of our result,
i.e. the discussion in Section~\ref{Context}.


\providecommand{\bysame}{\leavevmode\hbox to3em{\hrulefill}\thinspace}

\end{document}